\documentclass[sigconf]{acmart}
\pdfoutput=1

\usepackage{booktabs} 
\usepackage{tablefootnote}
\usepackage{balance}

\usepackage{amsmath}
\usepackage{amssymb}    
\newcommand{\sign}{\text{sgn}}

\DeclareMathOperator{\sgn}{sgn}

\begin{document}

\copyrightyear{} 
\acmYear{} 
\setcopyright{none=true}
\settopmatter{printacmref=false, printfolios=false}

\acmConference[EASE 2020]{Evaluation and Assessment in Software Engineering}{April 15--17, 2020}{Trondheim, Norway}

\acmPrice{}
\acmDOI{}
\acmISBN{}

\title[Assessing Software Defection Prediction Performance]{Assessing Software Defection Prediction Performance: Why Using the Matthews Correlation Coefficient Matters}

\author{Jingxiu Yao}
\affiliation{%
  \institution{Beihang University, China}
}
\email{yjx1080@126.com}

\author{Martin Shepperd}
\affiliation{%
  \institution{Brunel University London, UK}
}
\email{martin.shepperd@brunel.ac.uk}

\begin{abstract}
\emph{\textbf{Context}}: There is considerable diversity in the range and design of computational experiments to assess classifiers for software defect prediction.  This is particularly so, regarding the choice of classifier performance metrics.  Unfortunately some widely used metrics are known to be biased, in particular $F_1$.\newline
\emph{\textbf{Objective}}: We want to understand the extent to which the widespread use of the $F_1$ renders empirical results in software defect prediction unreliable.\newline
\emph{\textbf{Method}}: We searched for defect prediction studies that report both $F_1$ and the Matthews correlation coefficient (MCC).  This enabled us to determine the proportion of results that are consistent between both metrics and the proportion that change. \newline
\emph{\textbf{Results}}: Our systematic review identifies 8 studies comprising 4017 pairwise results.  Of these  results, the direction of the comparison changes in 23\% of the cases when the unbiased MCC metric is employed.  \newline
\emph{\textbf{Conclusion}}:  We find compelling reasons why the choice of classification performance metric matters, specifically the biased and misleading $F_1$ metric should be deprecated.
\end{abstract}


\maketitle
\section{Introduction}\label{Sec:Intro}
\noindent
A considerable body of empirical software engineering research is dedicated to predicting which are the defect-prone components in software systems \cite{Cata09,Hall12,Radj13,Malh15,Hoss17}.  The motivation is quite straightforward.  If we know which parts of the system are likely to be problematic, we can allocate testing resources accordingly. Moreover, more recent studies also take into account varying component test costs e.g., depending on size \cite{Mend10}.  Another important line of research is into the ability to use data from one software project to make predictions for a different project \cite{Zimm09}.  

Given this interest in software defect prediction, it is no surprise that there have been hundreds of studies that experimentally compare competing prediction systems (e.g., logistic regression, neural nets, support vector machines, etc.) over one or more software defect data sets.  The comparisons are made on the basis of classification performance.  However, there are many ways in which one can measure performance and no single metric has been adopted, although the $F_1$ measure or its components: precision and recall are very widely used \cite{Malh15,Hoss17}.\footnote{Some early studies used accuracy but this has been widely deprecated for some while, due to its inability to factor in the chance component arising from using the modal class as a prediction \cite{Soko06}.}  Unfortunately in recent years, statisticians have pointed out that this metric is problematic \cite{Powe11,Hern12} and that there are better alternatives such as the Matthews correlation coefficient (MCC) \cite{Bald00}.  The question then arises: does it matter; should we regard this as marginal statistical trivia or conversely, we should reject all defect prediction experiments based on unsound performance metrics?  

The problem of how to view results based on unsound metrics becomes all the more pressing, given the significant efforts now being deployed to aggregate experimental studies in software defect prediction.  It is well known that meta-analysis needs to filter out unreliable primary studies if the overall outcome is to be in itself reliable \cite{Khan96}.  Software defect prediction examples include the widely cited systematic reviews by Hall et al.~\cite{Hall12} and Hosseini et al.~\cite{Hoss17}.

The aim of this paper is to help assess how much confidence can we have in past software defect prediction research that relies on the $F_1$ performance metric to derive empirically-based conclusions?  We specifically focus on $F_1$ given it is known to be biased, particularly in the context of imbalanced data sets, i.e., when there is a low  (or very high) prevalence of the positive case, which for us are the defect-prone software components.  It is also in widespread use.

Our specific research questions are:
\begin{itemize}
\item RQ1: How much practical difference, i.e., conclusion instability \cite{Menz12} is there between the classification performance metrics $F_1$ and MCC?  Are they concordant?
\item RQ2: How does imbalance (the proportion of positive cases) impact differences between $F_1$ and MCC?
\item RQ3: How does the magnitude of differences in the classifier predictive performance impact differences between $F_1$ and MCC?
\end{itemize}

We believe these are important questions the defect prediction community needs to ask itself given the high proportion of past work that is based on $F_1$.  It also potentially allows us to identify if we can predict past results that are likely to be more or less unreliable.  Of course, for the future it should spur researchers to utilise alternative classification performance metrics for their experiments.  Or at least minimally, undertake full reporting so that meta-analysts and others can calculate metrics like MCC from the confusion matrices.

The remainder of the paper is organised as follows. The next section briefly reviews the diversity of classification performance metrics, specifically it contrasts the problematic $F_1$ metric with MCC and looks at experimental practices in software defect prediction.  Section~\ref{Sec:SysRev} describes the conduct and results of our systematic review to locate all relevant studies that present results both as $F_1$ and MCC. This provides us with the data to establish how concordant the two metrics are in practice.  This is followed by our analysis in Section~\ref{Sec:Res} of how results from the two different metrics compare.  Finally, in Section~\ref{Sec:Concl} we conclude with a discussion of the implications of our analysis and its limitations.

\section{Related Work}\label{Sec:Related}

\subsection{Classifier performance metrics}
\noindent
There is a wide range of metrics to assess the predictive performance of a classifier.  We restrict our discussion to the two-class problem, i.e., a software component is defect-prone (the positive class) or defect-free (the negative class).  This is because the vast majority of defect prediction research has adopted this stance \cite{Hall12}. Moreover, most performance metrics are based on the confusion matrix, which is shown in Table \ref{table_confusion_matrix}.  However, most metrics are, in theory, capable of generalisation to $k$-class prediction problems \cite{Ferr09}.  The confusion matrix contains the following counts: 
\begin{itemize}
\item true positives ($TP$) - defective components correctly classified as defective
\item true negatives ($TN$) - defect-free components correctly classified as defect-free
\item false positives ($FP$) - defective components incorrectly classified as defect-free
\item false negatives ($FN$) - defect-free components incorrectly classified as defective
\end{itemize}

\begin{table}
\renewcommand\arraystretch{1.5}
\centering
\setlength{\tabcolsep}{3pt}
\caption{The confusion matrix}
\label{table_confusion_matrix}
\begin{tabular}{llll}
\hline
 &  Actual Defective  & Actual Defect-free \\
\hline
Predicted Positive&
$TP$ &  $FP$\\
\hline
Predicted Negative&
$FN$ & $TN$\\
\hline
\end{tabular}
\end{table}

For our discussion regarding confusion matrices we use the following terminology:
\begin{enumerate}
\item Cardinality $n$ which is the total number of cases (i.e., $TP + FP + TN + FN$).
\item Defect density $d$ which is $TP + FN /n$ i.e., the total number of positive cases divided by all cases.
\item True positive rate ($TPR$), also referred to as sensitivity or recall is $TP/(TP + FN)$ which is the proportion of positive cases that are correctly considered as positive (defect-prone) as a proportion of all positive cases.
\item False positive rate ($FPR$) is defined as $FP/(FP+TN)$ which is the proportion of negative cases that are mistakenly considered as positive as a proportion of all negative cases.
\item $Precision$ is defined as $TP/(TP+FP)$ which is the proportion of correctly identified defect-prone moudles from all the cases classified as defective.
\item Accuracy is defined as $(TP+TN)/(TP+TN+FP+FN)$ which means the proportion of cases correctly classified to all cases.
\item $F_1$ is the harmonic mean of $TPR$ and $Precision$. It's based on F-measure which is defined as $(\beta^2 + 1)\times Precision\times TPR/(\beta^2\times Precision + TPR)$. Where $\beta$ is used to regulate the weight given to $TPR$. $F_1$ is one of the situation of F-measure when $\beta$ is set to 1 which indicates the weight of $TPR$ and $Precision$ is the same.
\item MCC is defined based on TP, TN, FP and FN, which includes all parts of the confusion matrix. Its range goes from -1 to +1 and higher values represent better performance. A value of +1 indicates a perfect prediction, -1 indicates a perverse prediction, and 0 indicates random predictions i.e., no classification value. 
\item The G-mean (GM) is calculated as the geometric mean of $TPR$ and $1-FPR$.
\end{enumerate}

\begin{table}
\renewcommand\arraystretch{1.5}
\centering
\setlength{\tabcolsep}{3pt}
\caption{Commonly used classification performance metrics}
\label{table_performance_metrics}
\begin{tabular}{llll}
\hline
Metrics&  
Definition&
Range&
Better\\
\hline

$TPR$&
$\frac{TP}{TP+FN}$&
[0,1]&
High\\
\hline
$FPR$&
$\frac{FP}{TN+FP}$&
[0,1]&
Low\\
\hline
$Precision$&
$\frac{TP}{TP+FP}$&
[0,1]&
High\\
\hline
$Accuracy$&
$\frac{TP+TN}{TP+TN+FP+FN}$&
[0,1]&
High\\
\hline
$F1$&
$\frac{2\times TPR\times Precision}{TPR+ Precision}$&
[0,1]&
High\\
\hline
$MCC$&
$\frac{TP\times TN- FP\times FN}{\sqrt{(TP+FP)(TP+FN)(TN+FP)(TN+FN)}}$&
[-1,1]&
High\\
\hline
$GM$&
$\sqrt{TPR\times(1-FPR)}$&
[0,1]&
High\\
\hline
$AUC$&
Area Under Curve (AUC) ROC chart&
[0,1]&
High\\
\hline
\end{tabular}
\end{table}

Table \ref{table_performance_metrics} summarises the performance metrics widely used in the previous studies.  Note that all metrics excepting AUC may computed from a single confusion matrix. The AUC differs in that it is based on a family of TPR and FPR values generated by changing the positive case acceptance threshold in small increments.  From this a frontier can be constructed which is known as a Receiver Operating Characteristic (ROC) curve \cite{Fawc06}.  AUC measures the area under this frontier where any value greater than 0.5 represents better than chance classification for the two-class case.  However, this metric refers to a family of possible classifiers rather than any specific classifier.  Thus, unless the ROC curve of classifier A strictly dominates classifier B we cannot make any remarks about our preference of A over B since, in practice, we can only deploy a single classifier.  For this reason we will not explore AUC further in this paper.  In passing we also note that  AUC has come under considerable criticism (see for example \cite{Hand09,Flac15}).

\subsection{A critique of $F_1$ and comparison with MCC} \label{Sec:F1vMCC}
\noindent
In this section we address the question of why is $F_1$ so problematic, particularly in the light of the fact that it continues to be widely deployed.  We then argue (actually repeat the arguments of others) that MCC (sometimes known as $\phi$ \cite{Warr08}) is a superior choice.

The $F_1$ is a specific instantiation ($\beta=1$) of the F-measure where:

\begin{equation*}
F_\beta = (1 + \beta^2) \cdot \frac{(\text{precision} \cdot \text{recall} )}{ (\beta^2 \cdot \text{precision})+ \text{recall} }
\end{equation*}

\noindent
It originates from the field of information retrieval and was proposed by van Rijsbergen \cite{vanR79} to measure the performance of information retrieval algorithms.  This problem domain is characterised by very large, indeed frequently unknowable, counts for TN.  Consider, for example, retrieving web pages.  Knowing the number of irrelevant pages correctly not retrieved, i.e., true negatives, will be both challenging and not very interesting.  It would run into hundreds of million, possibly more.  Unfortunately, applying $F_1$ in the completely different context of defect prediction is not equivalent.  A project manager or software engineer will be \emph{very} interested in knowing the number of correctly predicted defect-free (or negative) components.  These can have reduced, or in principle, zero (if the user is very trusting!), testing resources.  This calculation is major part of the rationale for software defect prediction.  Ignoring this important prediction outcome, results in $F_1$ being an unreliable guide to defect classification performance \cite{Powe11}.

The definition of $F_1$ (see Table~\ref{table_performance_metrics} is only based on TP, FP and FN since the metric is the harmonic mean of precision and recall.  This means the TNs are neglected.  The problem is compounded by much variation in the prevalence of defect-prone cases, but typically they are very much in the minority, in other words the data sets are imbalanced \cite{Sun09} unless some corrective procedure is undertaken \cite{Song19}.  

A second source of difficulty for $F_1$ is that it is a composite of two underlying measures: precision and recall.  This means two classifiers with very different properties might achieve the same metric score.  Users might have a strong preference for, say, high recall because they do not wish to miss fault-prone components, or alternatively for high precision in that they do not wish to squander valuable testing resources on defect-free components.  In theory one could weight the F-measure to follow user preferences, but it seems a near universal practice to have uniform weights hence $F_1$ sometimes termed the balanced F-score.  However, even some weighting procedure does not overcome the problem of it being possible to construct a particular metric value in multiple ways.

A third problem is that the metric is difficult to interpret other than zero is the worst case and unity the best case.  Specifically the chance component of the metric is unknown, unlike a correlation coefficient or AUC.  So, for example, it is hard to know what $F_1 = 0.25$ means.  Is it better than chance?  Is the classifier actually predicting?\footnote{In principle, we can compute the $F_1$ value associated with random estimation when the predictions follow the distribution of positive and negative cases such that $\text{TP/(TP+FN)} = \text{FP/(TP+FN)} \land \text{TN/(TN+FP)} = \text{FN/(TN+FP)}$.  Unfortunately, we cannot reason that any value of $F_1$ greater than chance is an improvement since it ignores all TNs so a classifier that is calamitously bad at predicting defect-free components and is worse than guessing could still achieve a `better' $F_1$ score.}  In contrast, a correlation coefficient equal to 0.25 means there is a small positive effect and that the classifier is doing better than chance.

We illustrate, these problems with an example.

\begin{equation*}
\left(\begin{array}{cc} \text{TP}=5 & \text{FP}=5 \\\text{FN}=5 &  \text{TN}=5\end{array}\right)   
\left(\begin{array}{cc} \text{TP}=5 & \text{FP}=5 \\\text{FN}=5 &  \text{TN}=85\end{array}\right)   
\end{equation*}



\noindent
Given the above two different confusion matrices, the lefthand matrix has $n=20$ and $d = 10/20 = 0.5$.  The distribution of predicted and actual classification is that of chance and hence no useful classification is taking place.  This is reflected in a correlation coefficient of $\text{MCC}=0.0$ and $F_1 = 0.5$.  A software tester would reject such a classifier as being no better than guessing.  Now contrast this matrix with the one on the right.  Here $n=100$ and $d = 10/100 = 0.1$, thus the data are highly imbalanced with the prevalence of positive cases being low.  Incidentally, this is a very typical situation for software defect prediction.  The difference between this new matrix and the previous example, is that the classifier is very effective at identifying defect-free cases and this is reflected in $\text{TN}=85$ and $\text{TNR}=85/(85+5) \approx 0.94$.   This  level of discrimination is likely to be valued in a practical setting since correctly identifying 90\%+ of the software components that don't need extra testing resources will be useful.  This is reflected in a correlation coefficient of $\text{MCC}\approx 0.44$ however, $F_1 = 0.5$ remains unchanged.  This is because it ignores the true negatives and so cannot distinguish between the two radically different classification situations as reflected in our example confusion matrices.

As a final remark, concerning this example, note the difficulty of understanding the meaning of $F_1 = 0.5$.  What proportion is due to chance odds? Can we be sure that meaningful classification is taking place?  Consider another confusion matrix where the accuracy is less than 0.5 (9/20).  In this case, $F_1 \approx 0.52$ and correctly $\text{MCC}\approx -0.11$ where the negative value connotes perverse performance!  

\begin{equation*}
\left(
\begin{array}{cc}\text{TP}=6 & \text{FP}=7 \\
\text{FN}=4 & \text{TN}=3\\ \end{array}
\right)
\end{equation*}

It is for this reason that many researchers using $F_1$ to assess classification performance restrict themselves to comparative analysis, i.e., is classifier A preferable to classifier B.  Yet even this is unsafe, because in the above three examples, the third one with a predictive performance of less than chance would be rated as best since $(F_1= 0.52) \succ (F_1=0.5)$.

For a more in depth evaluation and critique of $F_1$ as a performance classification metric see Sokolova et al.~\cite{Soko06,Soko09}, Warrens \cite{Warr08}, Hand \cite{Hand09}, Rahman et al.~\cite{Rahm12,Rahm13}  and Luque et al.~\cite{Luqu19}.  For an authoritative review of a wide range of classification metrics we recommend Powers \cite{Powe11}.

\subsection{The usage of classification performance metrics in software defect prediction}
\noindent
There are a wide range of metrics, however, widely used by software defect prediction researchers is the $F_1$ metric \cite{Malh15,Hoss17}.   This is unfortunate because, as we have explained in Section~\ref{Sec:F1vMCC}, it is known to be biased.

More specifically, Malhotra et al.~\cite{Malh15} found 17 out of 64 papers (1991-2013) used $F_1$ directly and a further 23 precision and 42 used recall, which are the constituent components of $F_1$ (see Table~\ref{table_performance_metrics}.   The more recent systematic review by Hosseini et al.~\cite{Hoss17} report that out of 30 studies (2006-2016) 11 use $F_1$ and 21 use precision and recall.

Corroborating these results, from our systematic review (2014-2019) we found 31 papers (where the content was available, in English and the application domain was software defect prediction).  Not all these papers satisfied other inclusion criteria (as discussed in Section~\ref{Sec:SysRev}) to enable us to compare $F_1$ with MCC, however we could make some judgement about the widespread usage of $F_1$.  We found that 29 out of 31 studies applied $F_1$ to compare the predictive performance of software defect classifiers.  Of course the nature of our search was to look for such papers, but it does indicate that many studies exist, even if we cannot make strong inferences about the proportion they represent.

So overall, it would probably seem that there is an increasing tendency to use $F_1$ in software defect studies.  Furthermore this then propagates through into meta-analyses which are often based on this metric \cite{Malh15,Hoss17} whilst other meta-analyses were obliged to discard data when researchers only reported results in terms of $F_1$ e.g., \cite{Shep14}.

\section{Systematic Review}\label{Sec:SysRev}
\noindent
In order to compare experimental results based on the $F1_1$ \emph{and} MCC metrics we needed to locate relevant, published research studies.  To do this we searched for published software defect prediction studies using a lightweight, systematic review strategy.  We deviated from the full-blown method described by Kitchenham et al.~\cite{Kitc15} principally in that there is no formal protocol and we did not make any quality assessment of the papers other than to require meaningful peer review and cross-validation.

The search was undertaken in November 2019.  We used the google scholar database based on the following query:\\

\texttt{"software defect"  classifier data F1 MCC}\\

\noindent
which located 53 results.  We then applied our inclusion criteria that a paper must satisfy.  
\begin{enumerate}
\item be published since 2014
\item be refereed i.e., we exclude the grey literature such as student dissertations and unpublished reports
\item have full content available 
\item be written in English
\item relate to software defect prediction
\item report both $F_1$ and MCC classification performance metrics for the \emph{same} comparisons of classifiers
\item make use of some form of cross-validation i.e., the results are based on unseen data
\item describe, new (i.e., not previously reported) results
\end{enumerate}

\noindent
The search results are summarised in Table \ref{Tab:Search} which yields a total of eight papers\footnote{An additional paper \cite{Mora17} is refereed, in scope and provides the necessary data but causes two types of difficulty.  Firstly, almost all of the comparisons between treatments result in ties, thus it would seem that the various treatments (algorithmic procedures for classification) are extremely similar.  Secondly, there are a large number (16) different treatments across 52 data sets potentially leading to more than 6000 pairwise comparisons.} for our analysis.  These 8 papers contain a total of 4017 pairwise comparisons between competing classifiers.  The number of results per paper varied considerably from 14 to 1512 with the median being 282.  In each case the comparisons are made twice, once with the $F_1$ metric and once with MCC.  Using this information we could next investigate the extent to which the two metrics are concordant (agree) and the extent to which they are discordant (lead to contradictory conclusions).

\begin{table}[htp]
\caption{Search inclusion criteria and paper counts}
\label{Tab:Search}
\begin{tabular}{|l|c|}
\hline
Criterion & Count \\
\hline
Total&53\\
Refereed&37\\
English&51\\
Available&46\\
Software defect prediction&28\\
Reports F1 and MCC&10\\
Cross validation&11\\
\hline
Included&8\\
\hline
\end{tabular}
\end{table}%

\begin{table*}[ht]
\caption{Summary of included software defect prediction studies}
\label{Tab:StudySummary}
\begin{tabular}{llrlllll}
  \hline
 Paper & Year & Type & Venue & $F_1$ & MCC & AUC & Data.sets \\ 
  \hline
Rodriguez et al.~\cite{Rodr14} & 2014 & C & EASE & Y & Y & Y & NASA MDP \\ 
Mausa Sarro Grbac \cite{Maus17} & 2017 & J & IST & Y & Y & N & PDE, Hadoop \\ 
Abaei et al.~\cite{Abae18} & 2018 & J & J. of KS Uni & Y & Y & N & NASA MDP \\ 
Tong Liu Wang \cite{Tong18} & 2018 & J & IST & Y & Y & Y & NASA MDP \\ 
Gong Jiang Jiang \cite{Gong19} & 2019 & J & SE\&P & Y & Y & Y & NASA MDP, Eclipse, SoftLab etc \\ 
Nezhad Shokouhi Majid \cite{Nezh19} & 2019 & J & J of Supercomputing & Y & Y & N & NASA  \\ 
Pan Lu Xu Gao \cite{Pan19} & 2019 & J & App Sci & Y & Y & N & PSC \\ 
Zhao Shang Zhao et al.~\cite{Zhao19} & 2019 & J & IEEE Access & Y & Y & Y & NASA MDP \\ 
\hline
\end{tabular}
\raggedright{{\footnotesize \\ NB C = conference; J = journal paper.}}
\end{table*}

Table \ref{Tab:StudySummary} provides an overview of the eight studies located by our systematic review and used as a source of results where we can compare the outcomes of using $F_1$ with MCC.  We can see that aside from Rodrigues et al.~\cite{Rodr14}, all the studies have been published since 2017 and they are all journal literature.  Although, we focus on comparing MCC with $F_1$, we note that AUC is also reported by 4 out of 8 studies.

\section{Results}\label{Sec:Res}
\subsection{Summary of the classification accuracy metrics}
\begin{figure}[htp]
\begin{center}
\includegraphics[width=\linewidth]{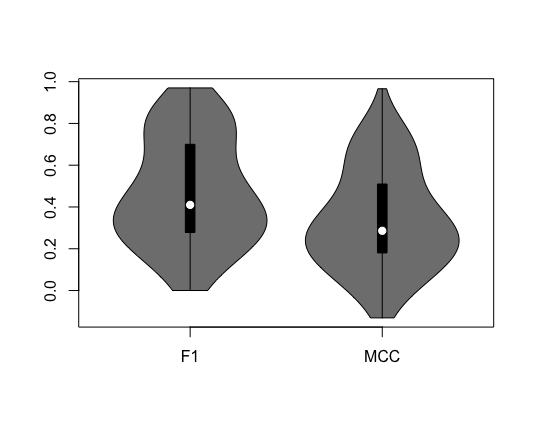}
\caption{Violin plots of $F_1$ and MCC}
\label{Fig:violin}
\end{center}
\end{figure}

\noindent
First, we summarise the accuracy metrics $F_1$ and MCC.  Recall that $F_1 = [0,1]$ whilst MCC = $[-1, 1]$ though in both cases, higher values imply superior classification performance. From Table \ref{Tab:MetricSummary} it is clear that both metrics vary considerably, from worse than random (or perverse classification) to near perfect.  $F_1$ in particular shows a bimodal distribution (see the violin plots in Fig.~\ref{Fig:violin}).

\begin{table}[ht]
\centering
\caption{Summary of the classification accuracy metrics}
\label{Tab:MetricSummary}
\begin{tabular}{|l|r|r|r|r|}
  \hline
 Metric &  Min &  Median  & Mean  & Max\\
 \hline
 $F_1$ & 0.00  &  0.41 & 0.47  &  0.97 \\
MCC & -0.13 & 0.29 & 0.34 &0.97\\   
 \hline  
\end{tabular}
\end{table}

\begin{table}[]
\centering
\caption{Odds ratio}
\label{Tab:odds_ratio}
\begin{tabular}{|l|l|l|l|}
\hline
 $F_1$ vs MCC          & upper third & lower third & total \\
\hline
change     & 186         & 455         & 641   \\
not change & 1153        & 884         & 2037  \\
\hline
total      & 1339        & 1339        & 2678 \\
\hline
\end{tabular}
\end{table}

\begin{figure*}[htp]
\begin{center}
\includegraphics[width=\linewidth]{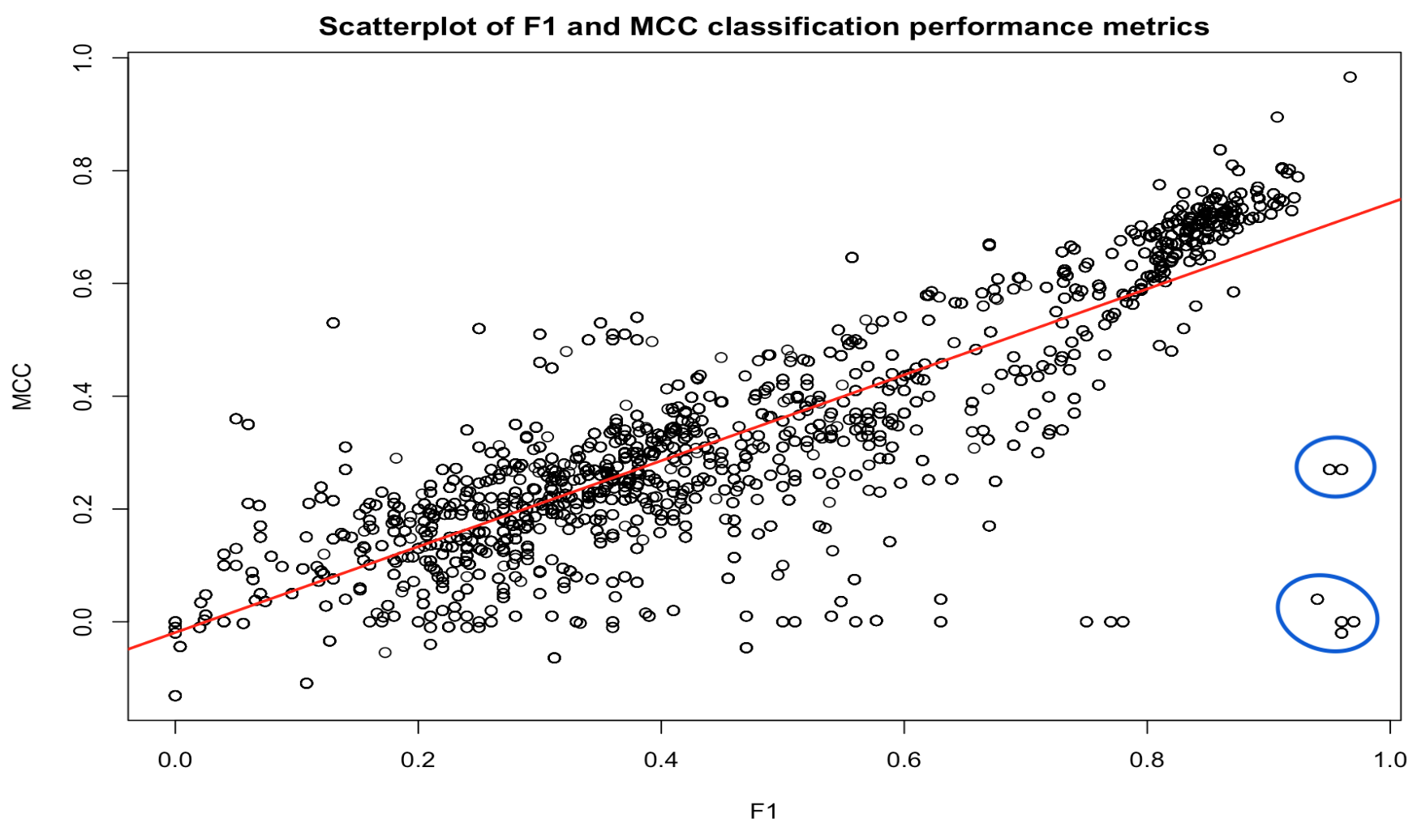}
\caption{Scatterplot of $F_1$ and MCC}
\label{Fig:scatter}
\end{center}
\end{figure*}

Since the metrics are measured on different scales we do not expect identical values, but would expect a monotonically increasing relationship.  Figure \ref{Fig:scatter} shows the relationship between the two classification accuracy metrics.  The red line shows the linear relationship between the two classification performance metrics.  Broadly, as one might hope, the relationship is positive thus as $F_1$ increases so does MCC.  However, there is a good deal of scatter and the Spearman correlation is $\rho \approx 0.82, n=8034$.  Some extreme outlier points are clearly visible and two clusters are circled for further investigation.

The two clusters are all drawn from the same primary study \cite{Pan19} and the same data set which is extremely imbalanced with, remarkably, in excess of 95\% of defective components from the Log4j project.  The reported values suggest near perfect $F_1$ scores and MCC values around or even below zero.  In other words there is no association between the predicted classes and the actual classes.  This is a little surprising and it is possible that the authors have treated the minority case (defect-free) as the positive case.  Nevertheless it highlights how misleading the $F_1$ metric can be in the face of highly imbalanced data.

\subsection{Summary of data set utilisation}
\noindent
In order to understand whether there is any relationship between data set imbalance and the concordance of the two classification metrics we also recorded data set information for each result.  In total 112 distinct data sets were employed, predominantly from the Promise archive or the NASA MDP data sets.  Different releases are viewed as different data sets.  Note that some primary studies explicitly used sub-samples of larger files in which case we counted these as separate unique data sets e.g., Zhao et al.~\cite{Zhao19}.  Where significant, and fully specified, data pre-processing is undertaken we regard these as separate data sets.  However, we accept the possibility that minor or undocumented pre-processing leads to slightly different variants of a data set.  In addition, the random allocation of cases to folds for cross-validation is likely to result in other subtle differences.

\begin{figure}[htp]
\begin{center}
\includegraphics[width=\linewidth]{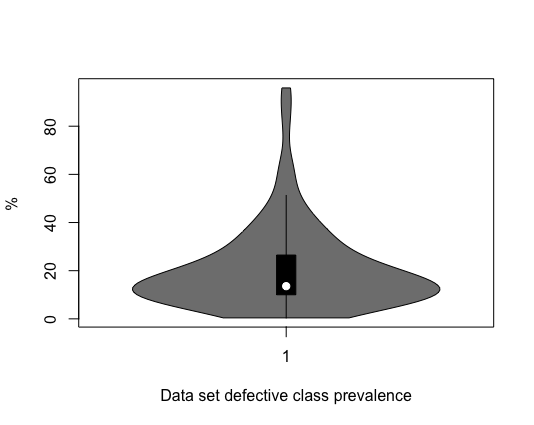}
\caption{Distribution of class-imbalance for software defect data sets}
\label{Fig:DataViolinPlot}
\end{center}
\end{figure}

The data sets vary considerably from 36 to 17186 cases (or software components).  The median size is 400 and mean is 1141 cases, indicating a small number of very large data sets.  In terms of imbalance, this ranges from 0.4\% to 95.6\% of positive (defect-prone components). Given it is well known that class-imbalance causes problems for classifiers this is quite striking.  We show the distribution as a violin plot in Fig.~\ref{Fig:DataViolinPlot} where it can be seen that the median imbalance level is about 13.5\%.  Note that a very high imbalance is equally problematic to a low imbalance rate.  Probably for this reason, researchers applied imbalanced learning or corrective procedures (e.g., under or over-sampling) to 35 out of 112 data sets. 

Having briefly summarised the raw data from the eight primary studies identified by the systematic review we now turn to our three research questions.

\subsection{RQ1: Are $F_1$ and MCC concordant?}\label{SubSec:RQ1}
\noindent
In order to answer this question, we break down the (generally many) results from each paper into a series of pairwise comparisons.  Formally speaking, $F_1$ and MCC are concordant iff: 
\begin{equation*}
\sgn(Y1_{F1} - Y2_{F1}) = \sgn(Y1_{MCC} - Y2_{MCC})
\end{equation*}
\noindent
where $Y1$ and $Y2$ are two classification results derived from the application of two treatments, which in our domain of enquiry would be competing classifiers applied to the \emph{same} data set. NB the $\sgn$ function is a mathematical function that extracts the sign of a real number.  

Therefore, if both $F_1$ and MCC find that treatment $Y2$ is to be preferred to $Y1$, indicated by a positive sign, then we would say they are concordant and from practical point of view it doesn't matter which metric is employed.  If they have differing signs we would say they are discordant and the conclusions disagree.  As a simple example, consider the situation where if we analyse our experimental results using $F_1$ we find, say, Naive Bayes is to be preferred to Logistic Regression.  However if we conduct the same experiment and analyse the results using MCC we find that on the contrary, Logistic Regression out-performs Naive Bayes then we say the results are discordant; we would make a different decision depending upon which performance metric is chosen.

Typically, papers report results in tables where the rows constitute different software systems represented by different data sets (these are the experimental units) and the columns represent different treatments for classifying the software such as logistic regression, random forest and so forth.  This is a natural way to organise computational experiments that are based on a repeated measures design \cite{Shad02} that is characterised by each treatment being applied to \emph{every} experimental unit.  

\begin{table}[htp]
\begin{center}
\caption{Example Defect Prediction Results}
\label{Tab:Example}
\begin{tabular}{|l|c|c|c|}
\hline
Data set & $T1$ & $T2$ & $T3$ \\
\hline
D1 & $Y_{11}$  & $Y_{12}$ & $Y_{13}$ \\
D2 & $Y_{21}$  & $Y_{22}$ & $Y_{23}$ \\ 
\hline
\end{tabular}
\end{center}
\raggedright{{\footnotesize NB D1 and D2 are data sets, T1, ..., T3 are different treatments (for us these will be different defect classifiers) and the Ys are the response variables (for us, these will be the classification performance measured by $F_1$ and MCC.}}
\end{table}%

To give a simple example, suppose a paper contains three different treatments (T1, T2 and T3) applied to two different software projects or data sets (D1 and D2). This might lead to a table similar to that given in Table \ref{Tab:Example}.  In this case, each $Y_{dt}$ will have two associated classification accuracy values, one as $F_1$ and one as MCC.  From the Table \ref{Tab:Example} we can extract the pairwise treatments, i.e., as detailed by Table \ref{Tab:PWexample}.

\begin{table}[htp]
\begin{center}
\caption{Example Defect Prediction Results}
\label{Tab:PWexample}
\begin{tabular}{ccc}
$ \sign(Y_{11} - Y_{12}) $ & $\sign(Y_{11} - Y_{13}) $ & $ \sign(Y_{12} - Y_{13}) $ \\
$ \sign(Y_{21} - Y_{22}) $ & $\sign(Y_{21} - Y_{23}) $ & $ \sign(Y_{22} - Y_{23}) $ \\
\end{tabular}
\end{center}
\end{table}%

Papers either report the different performance metrics side by side as columns in the same table, or in successive tables, but either way we can then compare the signs from the two metrics for each pairwise comparison. By analysing all the reported results in our 8 located papers we finish up with, as previously stated, 4017 pairwise comparisons where each $F_1$ result can be matched with an associated MCC result.  Then we compare whether the signs are equal to determine whether the metrics are concordant or not.

Such tables and our decomposition into pairwise comparisons form the basis of reasoning and forming conclusions regarding an experiment's results.  We adopt this approach for three reasons.  First, it enables us to unify experimental results --- presented by different papers in different styles --- into a single format for our analysis. Second, the notion of concordancy allows us to abstract away from the specifics of the individual purposes, hypotheses and results of each experiment and the means of making inferences.  Third, it aligns with the idea of using preference relations informed by experimental results to guide the decision-making of software engineers.

\begin{figure}[htp]
\begin{center}
\includegraphics[width=\linewidth]{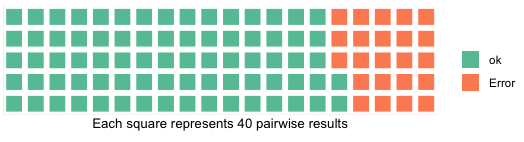}
\caption{Proportion of incorrect experimental results}
\label{Fig:waffle}
\end{center}
\end{figure}

In total we find that \textbf{23\% (927/4017)} of comparisons or conclusions are discordant between the $F_1$ metric and MCC.  This is shown graphically by Fig.~\ref{Fig:waffle}.  Given that the $F_1$ metric is known to be biased, this means that an experimenter could be misled (i.e., accepting the wrong preference relation and thinking the first classification is to be preferred to the second, when in fact it is the other way around) almost one in four times.

\subsection{RQ2: How does data set imbalance impact differences between $F_1$ and MCC?}

\begin{figure*}[htp]
\begin{center}
\includegraphics[width=\linewidth]{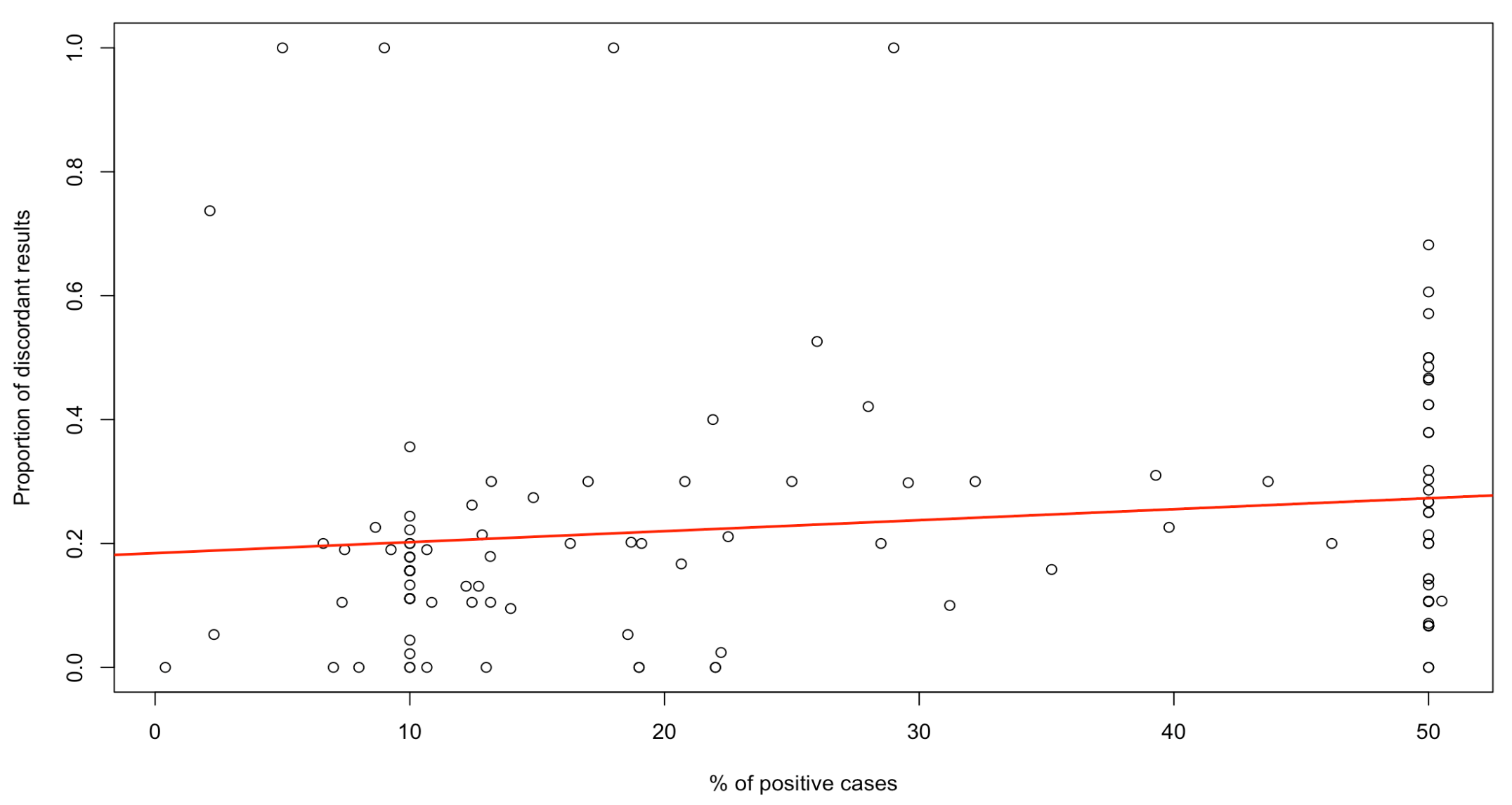}
\caption{Scatterplot of defect data set (post-processed) imbalance level and the proportion of discordant results}
\label{Fig:PostImbalScatter}
\end{center}
\end{figure*}

Next we consider whether there is any pattern to this discordancy, since we might hypothesise that highly imbalanced data sets could cause more problems for $F_1$ since it is a biased metric.  We examine the relationship between the post-processed imbalance rate since this is what experimenters actually use.  Fig.~\ref{Fig:PostImbalScatter} shows relatively little relationship.  We also find little association between post-processed imbalance and the probability of a result changing (Spearman's $\rho=0.24, n=112$).  What is more, the direction of the relationship is the opposite of what one might expect in that data sets with closer to 50\% defect-prone software components appear more likely to have discordant results.  Consequently we move onto our third research question to see whether the size of the effect moderates the impact of data set imbalance.

\subsection{RQ3: How does the magnitude of difference in the classifier performance impact discordancy?}

The underlying hypothesis for this research question is that when the difference between classifier performance is relatively trivial then discordancy between $F_1$ and MCC metrics is more likely.  Ideally we would couch this question in terms of a standardised effect size \cite{Elli10}.  Unfortunately such measures cannot be easily constructed given the absence of reported standard errors or other statistics of dispersion.  However, given the scale of the $F_1$ performance metric is defined as $[0, 1]$, we do have some basis for comparison and will examine the absolute differences, between pairs of $F_1$ metrics.


\begin{figure}[htp]
\begin{center}
\includegraphics[width=\linewidth]{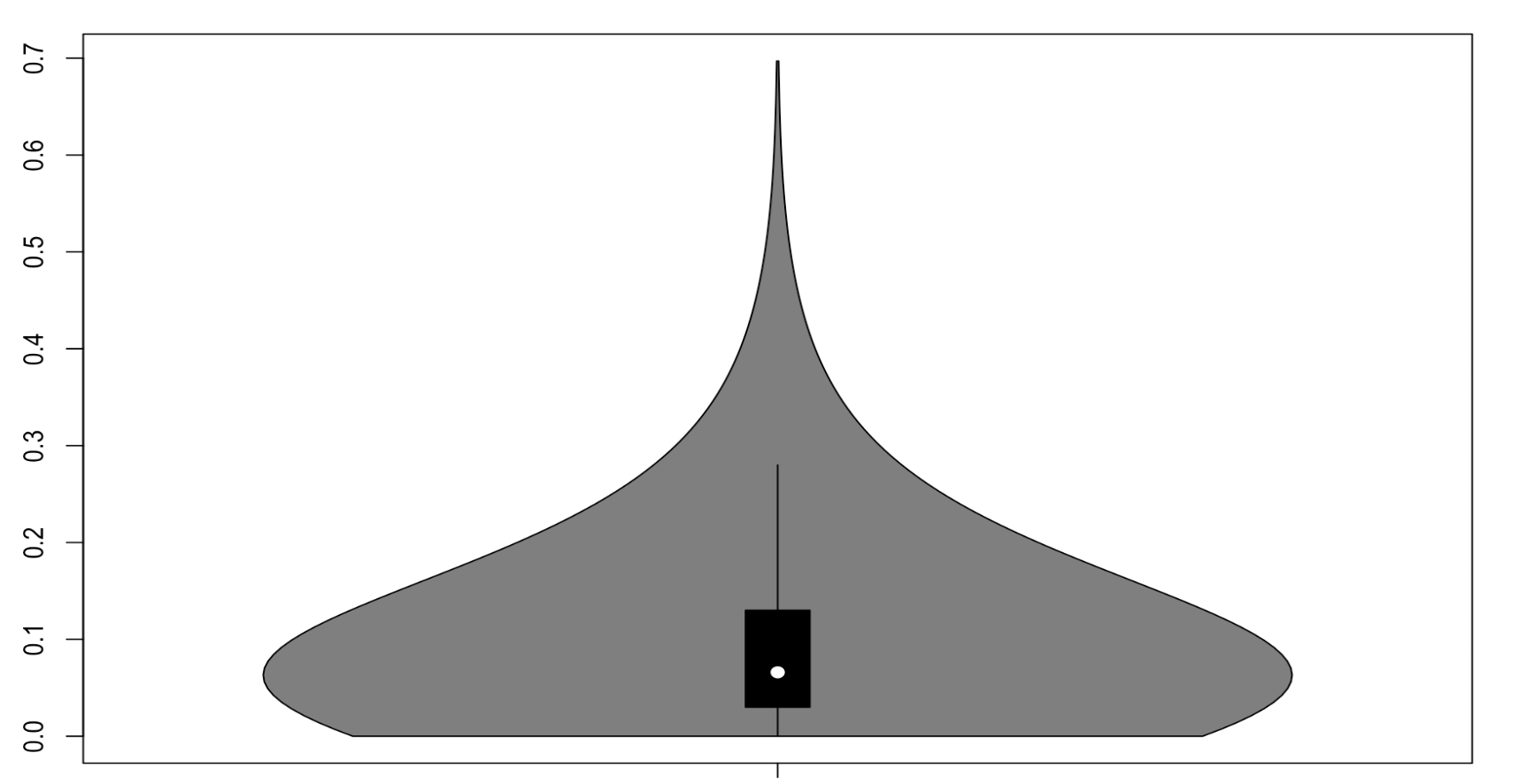}
\caption{Absolute difference between treatments T1 and T2 measured by $F_1$}
\label{Fig:F1diffViolin}
\end{center}
\end{figure}
   
Fig.~\ref{Fig:F1diffViolin} shows the distribution of `effects' captured as the absolute difference between the two treatments for each pairwise comparison.  This suggests that the majority of comparisons between classifiers yield comparatively small differences as captured by the $F_1$ metric with a median of 0.066.

\begin{figure}[htp]
\begin{center}
\includegraphics[width=\linewidth]{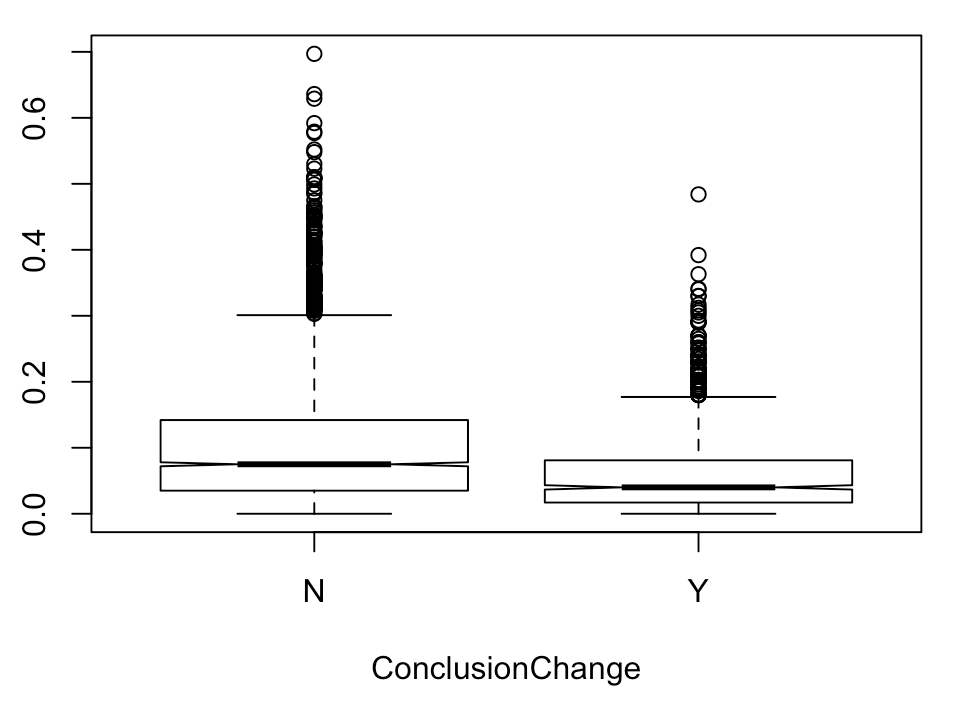}
\caption{Comparison of the absolute difference between treatments T1 and T2 measured by $F_1$ for concordant and discordant results}
\label{Fig:F1diffBoxplot}
\end{center}
\end{figure}

If we split the absolute $F_1$ differences by whether there is a conclusion change (i.e., discordancy) or not, we observe typically greater `effects' (see Fig.~\ref{Fig:F1diffBoxplot}) where there is no change.  This might be expected.  The stronger the effect, as captured by the absolute difference of the two $F_1$ metric values, the less likely it will be different from the equivalent pair of MCC metrics.  

We also investigate this phenomenon using a variant of the median split, where the pairwise comparisons are ranked by absolute $F_1$ difference and split into thirds.  The middle third is discarded and then we construct the odds ratio of there being a discordancy with the MCC metric between the upper and lower third.  We represent the raw scores as Table \ref{Tab:Odds} and compute the odds ratio of a pairwise conclusion being discordant from the lower third compared with the upper third as 4.36 with a confidence interval (3.53, 5.39).  This suggests that the absolute difference in $F_1$ scores for a pairwise comparison of classifiers is some indicator of the likelihood that the unbiased MCC metric confirms this finding.  In other words, using $F_1$ is more problematic when there are only small differences between the classifiers being compared.

\begin{table}[htp]
\begin{center}
\caption{Comparison of discordant pairwise comparisons by upper and lower thirds of absolute comparison size for $F_1$}
\label{Tab:Odds}
\begin{tabular}{|l|c|c|c|}
\hline
Concordant? & Upper third & Lower third & Total \\
\hline
No & 186 & 455 & 641 \\
Yes	 & 1153 & 884 & 2037 \\
\hline
Total & 1339 & 1339 & 2678 \\
\hline
\end{tabular}
\end{center}
\end{table}%

\section{Discussion and Conclusions}\label{Sec:Concl}
\noindent
In this study we have sought to understand the impact of the widespread use of the problematic, classification performance metric $F_1$.  To do this we have conducted a systematic search to find primary experimental studies in the domain of software defect prediction.  Specifically, we sought experiments that have reported both $F_1$ (a biased metric) and the Matthews correlation coefficient (a preferred metric).  We have then decomposed the results into a series of pairwise comparisons, of Classifier 1 versus Classifier 2.  Each comparison can be thought of as a preference relation, for instance that Classifier 1 is to be preferred to Classifier 2.  We can then compare each comparison using $F_1$ with MCC.  We found eight usable primary studies that contained a total of 4017 pairwise comparisons based on 112 data sets.\footnote{Our data are made available at {https://zenodo.org/deposit/3581263}.}  Given that there are concerns about $F_1$, we would be reassured if the comparisons using MCC is concordant with the $F_1$ comparisons, i.e., the sign function is the same.

More generally, we can interpret the level of concordance as a guide as to how reliable we can view research results emanating from the experiments that use $F_1$ as the response variable.  It can help us determine if the problems associated with $F_1$ are essentially academic, or whether they undermine our ability to trust such findings.

Whilst we did not find the majority of $F_1$ results discordant from an MCC analysis, we still consider 23\% to be very worrying.  This means almost a quarter of published results based on the flawed $F_1$ are likely to be incorrect.  By incorrect we mean that the direction of the comparison is in error.  Minimally we suggest this implies that such experimental results need to be treated with caution.

Finally, we wish to stress, our analysis implies no criticism of any of the eight primary studies we have used for our analysis.    Each study is refereed and has undertaken rigorous empirical analysis.  It is greatly to the credit of these researchers that they have reported a wide range of metrics and without this, our analysis would not have been possible.

So what lessons can we extract?

\begin{enumerate}
\item We should stop using biased performance metrics and, in terms of this study, specifically we should stop using the $F_1$ metric inappropriately.
\item Open science and full reporting is essential. Reporting the confusion matrix enables other researchers to compute a wide range of alternative metrics.  Even if the original researchers are disinclined to use metrics such as MCC, this should not prevent other researchers re-analysing and, if need be, re-interpreting published experiments.
\item Where readers must depend upon experimental results based on $F_1$ due to incomplete reporting and lack of alternatives, considerable caution should be deployed since the base rate odds of a result being in error is almost one in four.  Such a choice is most risky when the effect sizes are small.
\item We recommend that meta-analyses should avoid, wherever possible, primary studies solely based on the use of $F_1$.  Minimally, sensitivity analysis should be undertaken to compare the analysis with and without such primary studies.
\end{enumerate}
 
To return to the fundamental question, does it matter which classification performance metrics we use in our software defect prediction experiments?  Unfortunately the answer is: yes, very much so.  Clearly, there is more work to be done.  In particular, it would be interesting to examine the grey literature and also the use of classifiers in problem domains beyond software defect prediction.

\subsection{Threats to validity}
\noindent
As always there are a number of threats to validity.  In terms of \emph{Internal validity} this refers to whether our analysis captures the actual constructs in question.  
\begin{itemize}
\item Is the $\sgn$ function an appropriate way to capture a conclusion change?  It has the merit of being general and therefore not tied to any particular experiment or type of analysis such as null hypothesis significance testing.  It certainly captures a change in direction of the effect.  It does not, however, capture the magnitude of the effect. In the end our view is that finding so many changes in effect direction when comparing $F_1$ results with the more robust MCC is emblematic of underlying problems.
\item Imbalance is not precisely known.  This is because sampling cases into folds for cross-validation is a stochastic process.  Also exact details of any data pre-processing are not always fully reported.  These issues are likely to be more significant for the smaller data sets.  Nevertheless, the defect densities for each data set give a reasonable approximation of the imbalance, and the lack of a relationship between imbalance and concordance is sufficiently strong to not be materially impacted by more exact imbalance measurement.
\end{itemize}

For\emph{ external validity} threats relating to generalisability, we see the following issue.

Although we searched systematically, our search was quite tightly defined so we did exclude relevant experiments.\footnote{Following the suggestion one of the reviewers, we extended the search to consider other synonyms for F1 such as F-measure.  This yielded an additional eight papers containing a further $\sim4200$ pairwise comparisons.  Unfortunately the timescales do not permit us to complete the analysis, however, once it is completed we will place the new results on our open repository at  {https://zenodo.org/deposit/3581263}.}  We could have used additional queries and used other bibliographic databases. We could also have examined the grey literature. In mitigation, we believe we have some personal domain knowledge and we focused on recent papers (primarily because use of the Matthews correlation coefficient is a relatively recent phenomenon) and are not aware of any papers that have been missed. It has also been our intention to focus on high quality papers.  Informally, it has been our impression that there has been a high degree of scholarship and all papers passed our inclusion criterion of using cross-validation procedures.

Also, we only found eight experiments so is this a sufficiently large sample?  Of course a larger sample would be preferable but we are constrained by what is available. In addition, the eight papers contain over 4000 individual pairwise results which is a substantial body of data to analyse.  In the future this threat might also be tackled by exploring other problem domains.

\section*{Acknowledgements}
Jingxiu Yao wishes to acknowledge the support of the China Scholarship Council.
\bibliographystyle{ACM-Reference-Format}
\balance
\bibliography{JingXiuEASErefs}



\end{document}